\documentclass[12pt,twoside]{article}

\usepackage[margin=1in]{geometry}
\usepackage{amsmath,amssymb,amsfonts,amsthm}
\usepackage{graphicx}
\usepackage{hyperref}
\usepackage{listings}
\usepackage{physics} 
\usepackage{verbatim}
\usepackage{float}
\usepackage{parskip}

\hypersetup{
    colorlinks=true,
    linkcolor=blue,
    filecolor=magenta,      
    urlcolor=cyan,
}

\usepackage{xcolor}
\definecolor{codegreen}{rgb}{0,0.6,0}
\definecolor{codegray}{rgb}{0.5,0.5,0.5}
\definecolor{codepurple}{rgb}{0.58,0,0.82}
\definecolor{backcolour}{rgb}{0.95,0.95,0.92}

\lstdefinestyle{mystyle}{
    backgroundcolor=\color{backcolour},   
    commentstyle=\color{codegreen},
    keywordstyle=\color{magenta},
    numberstyle=\tiny\color{codegray},
    stringstyle=\color{codepurple},
    basicstyle=\ttfamily\footnotesize,
    breakatwhitespace=false,         
    breaklines=true,                 
    captionpos=b,                    
    keepspaces=true,                 
    numbers=left,                    
    numbersep=5pt,                  
    showspaces=false,                
    showstringspaces=false,
    showtabs=false,                  
    tabsize=2
}
\title{Quantum Cellular Automaton Model for the Generalized Dirac Equation}
\author{Xingyou Song\thanks{Email: xsong@berkeley.edu. This work was conducted during a research internship in the UCSD Quantum Computing Group in the summer of 2013.}}
\date{}

\begin{document}

\maketitle

\begin{abstract}
This study presents a unitary quantum cellular automaton (QCA) that, in the continuum limit, converges to the (1+1)-dimensional Generalized Dirac Equation (GDE). We outline the construction of the unitary, discrete-time evolution and derive the model's exact dispersion relation from its eigenvalue spectrum, showing it possesses the correct relativistic limit. We then explore its dynamics through numerical simulations. While the total probability density is shown to be insensitive to the chiral angle parameter $(\rho)$ of the GDE, we demonstrate that this parameter has a profound and directly observable effect on the particle's local spin polarization. By measuring this quantity, we reveal the hidden internal dynamics of the spinor, providing a clear, visual confirmation of the physical consequences of the chiral mass term.
\end{abstract}


\section{Introduction}

The simulation of quantum systems is a foundational challenge in physics and computer science. As Richard Feynman famously noted, classical computers face exponential difficulty in simulating quantum mechanics, suggesting that a computer built on quantum principles would be an intrinsically better tool for the task \cite{feynman1982}. This insight has motivated the development of quantum algorithms for simulating physical phenomena. Among the most promising paradigms for this are Quantum Cellular Automata (QCA) and Quantum Random Walks (QRWs).

QRWs, the quantum mechanical analogue of classical random walks, leverage superposition and interference to exhibit dynamics that are profoundly different from their classical counterparts \cite{kempe2003, venegas2012}. This has led to significant speedups in certain computational problems, such as search algorithms \cite{childs2004}. Beyond computation, the discrete, local-update structure of QRWs makes them natural frameworks for modeling the evolution of quantum systems governed by relativistic wave equations.

The Dirac equation \cite{dirac1928}, which describes the motion of spin-1/2 fermions, has been a key target for such models. Several studies have successfully formulated QRWs that converge to the (1+1)-dimensional Dirac equation in the continuum limit \cite{meyer1996, arrighi2013}. A critical requirement for any such physical simulation is that the evolution operator must be unitary. This paper constructs a unitary QCA for the Generalized Dirac Equation (GDE), a form of the Dirac equation with a chiral mass term, and presents direct simulations of its dynamics, revealing the observable consequences of this generalization.

\section{Formalism of Quantum Walks on a Lattice}
\label{formalism}

\subsection{The Hilbert Space of a Lattice Fermion}
We consider a quantum particle on a discrete one-dimensional lattice with $n$ sites, labeled by the integer position index $j \in \{1, 2, ..., n\}$. The simulation proceeds in discrete time steps labeled by the integer index $t$. The physical position is given by $x=j\varepsilon$ and physical time by $T_{phys}=t\varepsilon$, where $\varepsilon$ is the fundamental lattice spacing. Since the Dirac equation describes a spin-1/2 particle, the particle at each site also possesses an internal spin degree of freedom, which we label spin-up ($\ket{+}$) and spin-down ($\ket{-}$).

The total state of the particle is therefore described within a composite Hilbert space, $\mathcal{H} = \mathcal{H}_{\text{position}} \otimes \mathcal{H}_{\text{spin}}$. The basis for this space is the set of states $\{\ket{j, s}\}$. A general quantum state $\ket{\psi}$ at time step $t$ is a superposition:
\begin{equation}
\ket{\psi(t)} = \sum_{j=1}^{n} (\psi_{+}(j,t)\ket{j,+} + \psi_{-}(j,t)\ket{j,-}),
\end{equation}
where $\psi_{\pm}(j,t) \in \mathbb{C}$ are the complex probability amplitudes. The normalization condition requires the total probability to be one: $\sum_{j,s} |\psi_{s}(j,t)|^2 = 1$.

\subsection{Unitary Evolution via Operator Splitting}
The evolution of the state over a single discrete time step is governed by a global unitary operator, $M$. For a quantum walk that models the Dirac equation, this evolution can be broken down into two distinct, simpler unitary operations, a method known as operator splitting or the Trotter-Suzuki decomposition \cite{trotter1959}. The total evolution for one step is given by:
\begin{equation}
    \ket{\psi(t+1)} = M \ket{\psi(t)} = S \cdot C \ket{\psi(t)}
\end{equation}
where $C$ is the ``Coin" or ``Kick" operator, and $S$ is the ``Shift" or ``Stream" operator.

\paragraph{The Coin Operator (C):} This is a local operation that acts only on the internal spin state at each lattice site, without changing the particle's position. It can be represented by a 2x2 unitary matrix, $U_C$, that is applied to the spinor $(\psi_+, \psi_-)$ at every site. This step is responsible for mixing the spin components and implementing the effects of mass.
\begin{equation}
\ket{\psi'(t)} = C \ket{\psi(t)} \implies \begin{pmatrix} \psi'_{+}(j,t) \\ \psi'_{-}(j,t) \end{pmatrix} = U_C \begin{pmatrix} \psi_{+}(j,t) \\ \psi_{-}(j,t) \end{pmatrix} \quad \forall j
\end{equation}

\paragraph{The Shift Operator (S):} This operation moves the particle to adjacent lattice sites, with the direction of movement conditioned on the particle's spin state. For the Dirac equation, the shift operator typically moves the spin-up component one step to the right and the spin-down component one step to the left.
\begin{equation}
\ket{\psi(t+1)} = S \ket{\psi'(t)} \implies \psi_{+}(j, t+1) = \psi'_{+}(j-1, t), \quad \psi_{-}(j, t+1) = \psi'_{-}(j+1, t) \quad \forall j
\end{equation}
This conditional shift is what creates the propagation and the characteristic ``light cone" structure. The combination of the local spin rotation (Coin) followed by the spin-dependent movement (Shift) is a standard and powerful method for constructing unitary QCA models of relativistic particles.

\subsection{Observables}
To analyze the simulation, we measure physical observables. The most fundamental is the \textbf{probability density} at each site:
\begin{equation}
P(j,t) = |\psi_+(j,t)|^2 + |\psi_-(j,t)|^2
\end{equation}
To probe the internal dynamics of the spinor, we also measure the \textbf{spin polarization}, which is the expectation value of the Pauli $\sigma_3$ operator \cite{pauli1927}. It measures the local difference between spin-up and spin-down probability:
\begin{equation}
\expval{\sigma_3}(j,t) = |\psi_+(j,t)|^2 - |\psi_-(j,t)|^2
\end{equation}
Plotting these two quantities reveals different aspects of the particle's evolution.

\section{Generalized Dirac Equation}

\subsection{From Schrödinger to Dirac: The Need for a Relativistic Equation}
The Schrödinger equation is the cornerstone of non-relativistic quantum mechanics, but it is fundamentally incompatible with special relativity because it treats time (first-order derivative) and space (second-order derivative) on unequal footing. The first attempt to create a relativistic quantum wave equation was the Klein-Gordon equation \cite{klein1926, gordon1926}, which can be derived directly from the classic relativistic energy-momentum relation, $E^2 = p^2c^2 + m^2c^4$ \cite{einstein1905_relativity}. Using the quantum operator substitutions $E \to i\hbar\partial_t$ and $p \to -i\hbar\nabla$, and setting $\hbar=c=1$, we get:
\begin{equation}
(-\partial_t^2 + \nabla^2)\psi = m^2\psi \quad \implies \quad (\partial_t^2 - \nabla^2 + m^2)\psi = 0
\end{equation}
While correct, this second-order equation historically presented issues with defining a positive-definite probability density. Dirac's brilliant insight was to search for a new equation that was first-order in both space and time, analogous to the Schrödinger equation, but fully Lorentz-covariant. He proposed an approach of the form $(i\partial_t + ...)\psi = m\psi$, demanding that applying the operator twice would recover the Klein-Gordon equation. This ``square root" of the Klein-Gordon operator could not be achieved with scalar coefficients, forcing the introduction of matrix-valued objects: the gamma matrices.

\subsection{The Standard Generalized Dirac Equation in 1+1 Dimensions}
In a (1+1)-dimensional spacetime, the role of the gamma matrices can be fulfilled by the 2x2 Pauli matrices. A standard representation (e.g., $\gamma^0 = \sigma_1, \gamma^1 = -i\sigma_2$) leads to the Dirac equation. To this, we can add a generalized, or ``chiral", mass term, which is also compatible with the underlying relativistic structure. This term includes a pseudoscalar component parameterized by an angle $\rho$, leading to the Generalized Dirac Equation (GDE):
\begin{equation}
(i\sigma_1 \partial_t - \sigma_2 \partial_x)\psi = m(\cos\rho \cdot I + i\sin\rho \cdot \sigma_3)\psi
\label{eq:standard_gde}
\end{equation}
This equation is the central object of our simulation. It is physically well-posed because it satisfies Dirac's original requirement, recovering the Klein-Gordon equation when its operator is applied twice.

\subsection{Discretization and Unitary Evolution}
To construct a QCA that converges to Eq. (\ref{eq:standard_gde}), we must define a discrete update rule on a spacetime grid with step size $\varepsilon$. A general local rule would define the state at site $j$ and time step $t+1$ as a linear combination of the states at the neighboring sites $j-1, j,$ and $j+1$ at time step $t$. The key challenge is to find a set of coefficients for this combination that not only approximates the GDE but also forms a globally unitary evolution.

The operator-splitting method provides a powerful and standard method to guarantee unitarity. The specific one-step update rule for our model emerges directly from the composition of the Coin and Shift operators, $M=S \cdot C$. By substituting the Coin and Shift operations, we arrive at the complete single-step evolution rule for our QCA:
\begin{align}
\psi_{+}(j, t+1) &= \cos(R) \cdot \psi_{+}(j-1, t) - i e^{-i\rho} \sin(R) \cdot \psi_{-}(j-1, t) \label{eq:update_plus_final} \\
\psi_{-}(j, t+1) &= \cos(R) \cdot \psi_{-}(j+1, t) - i e^{i\rho} \sin(R) \cdot \psi_{+}(j+1, t) \label{eq:update_minus_final}
\end{align}
where $R=m\varepsilon$. These two equations, which are unitary by construction, form the exact algorithm used to generate all simulations in this paper.

\subsection{Eigenvalue Spectrum and the Dispersion Relation}
This analysis provides a deep, analytical confirmation that our QCA correctly captures the physics of the Dirac equation. The ``fingerprint" of a relativistic system is its dispersion relation, which connects energy $E$ and momentum $p$. For the continuous Dirac equation, this is the famous relation $E^2 = p^2 + m^2$. We can derive the dispersion relation for our discrete QCA and show that it converges to this correct limit.

Because the QCA is translationally invariant, its eigenvectors are momentum eigenstates (plane waves). The evolution in Fourier space is governed by a 2x2 matrix $M_k$, and its eigenvalues $\lambda_k = e^{-i\omega_k \varepsilon}$, define the quasi-energy $\omega_k$ for each momentum mode $k$. A straightforward calculation yields the exact dispersion relation for our QCA:
\begin{equation}
\cos(\omega_k \varepsilon) = \cos(m\varepsilon)\cos(k\varepsilon)
\end{equation}
To check the continuum limit, we take $\varepsilon \to 0$. Taylor expanding both sides to second order gives:
\begin{equation}
1 - \frac{(\omega_k \varepsilon)^2}{2} + \dots \approx \left(1-\frac{(m\varepsilon)^2}{2}\right)\left(1-\frac{(k\varepsilon)^2}{2}\right) \approx 1 - \frac{(m\varepsilon)^2}{2} - \frac{(k\varepsilon)^2}{2}
\end{equation}
Simplifying, we arrive at:
\begin{equation}
\omega_k^2 \approx m^2 + k^2
\end{equation}
This is precisely the relativistic energy-momentum relation, confirming that our discrete QCA possesses the correct relativistic dynamics in the continuum limit.

\subsection{Path Integral Interpretation and Combinatorics of the Walk}
\label{sec:path_integral}
Beyond the matrix formalism, the QCA evolution permits a powerful physical interpretation as a path integral on the spacetime lattice. This perspective provides a deep combinatorial origin for the interference patterns observed in our simulations.

The amplitude of the particle at a final position $j$ after $t$ time steps is the coherent sum of the amplitudes of all possible paths the particle could have taken. A ``path" on our lattice is a specific sequence of $t$ discrete steps, where each step is either to the left or to the right. For a particle starting at the origin ($j=0$) to arrive at site $j$ after $t$ steps, it must have taken a specific number of right steps ($N_R$) and left steps ($N_L$):
\begin{equation}
N_R = \frac{t+j}{2}, \quad N_L = \frac{t-j}{2}
\end{equation}
At each of the $t$ vertices along any given path, our Coin operator $U_C$ acts as a local ``scattering matrix", providing the complex amplitudes for the particle to either continue or reverse direction. The total amplitude for a specific path is the product of the $t$ matrix elements chosen at each step.

For a particle starting at the origin with spin-up, the total amplitude to arrive at $(j,t)$ with spin-up is given by an exact combinatorial sum over all possible numbers of spin flips ($2k$):
\begin{equation}
\psi_{+}(j,t) = \sum_{k=0}^{\min(N_R, N_L)} \binom{N_R}{k} \binom{N_L}{k} (\cos (m\varepsilon))^{t-2k} (- \sin^2 (m\varepsilon))^k
\label{eq:combinatorial_sum}
\end{equation}
The intricate, wavy patterns of the \textit{Zitterbewegung} are a direct visualization of the massive quantum interference between the terms in this sum. A detailed derivation provided in Appendix A shows that in the continuum limit, this discrete sum converges to the propagator of the continuous Dirac equation, which is described by Bessel functions. This confirms from a third, independent perspective that our QCA is a correct and powerful model of the Dirac equation.

\paragraph{The Continuum Limit of the Path Integral.}
To connect our discrete combinatorial sum to the continuous Dirac equation, we take the continuum limit. We set the physical position and time as $x=j\varepsilon$ and $T_{phys}=t\varepsilon$, and take the limit $\varepsilon \to 0$ while keeping $x, T_{phys},$ and $m$ fixed. In this limit, the number of steps $t$ and the number of right/left moves $N_R, N_L$ go to infinity. Furthermore, the mass parameter $R=m\varepsilon$ becomes infinitesimal. The process of taking this limit, as established in the foundational work of Feynman and Hibbs on path integrals \cite{feynmanhibbs1965}, is outlined in Appendix \ref{appendix:path_integral}. The result is that the propagator is described by the Bessel function:
\begin{equation}
\psi(x,T_{phys}) \propto J_0(m\sqrt{T_{phys}^2-x^2})
\end{equation}
where $J_0(z) = \sum_{k=0}^{\infty} \frac{(-1)^k}{(k!)^2} \left( \frac{z}{2} \right)^{2k}$. The argument of the Bessel function, $m\sqrt{T_{phys}^2-x^2} = m\tau$, is the product of the particle's mass and the relativistic proper time. The emergence of this Lorentz-invariant quantity from a discrete model is a profound feature of these systems. Furthermore, the oscillatory nature of the Bessel function analytically explains the persistent, wavy interference fringes of the \textit{Zitterbewegung} seen in our simulations below.

\section{Numerical Simulations}
\subsection{Varying Mass}
To investigate the dynamics of the Generalized Dirac Equation, we performed direct numerical simulations of the corresponding unitary quantum walk for a single particle. We explored the evolution in two distinct regimes: the initial, short-time propagation that approximates behavior in open space, and the long-time evolution on a finite lattice where periodic boundary conditions become significant. Note that any value of $\rho$ does not change the overall probability density $P(j,t) = |\psi_+(j,t)|^2 + |\psi_-(j,t)|^2$, and thus we perform all simulations with a fixed chiral angle of e.g. $\rho=0$ to focus on the role of the mass parameter $R$.

Figure \ref{fig:dynamics} (Top) illustrates the effect of the mass parameter on the total probability density, $\abs{\psi}^2$. This scenario cleanly illustrates the fundamental properties of a relativistic particle. The evolution begins from a single site and propagates outwards, forming a ``light cone." The complex interference pattern inside the cone, known as \textit{Zitterbewegung}, is clearly visible, arising from the interplay between the spinor's different energy components. As the mass parameter $R$ is increased from left to right $R \in \{0.20, 0.80, 1.20\}$, the propagation speed of the wave packet slows, and the particle becomes more localized, as expected from relativistic theory.

Figure \ref{fig:dynamics} (Bottom) shows the long-time evolution of the same systems on a smaller, periodic lattice. Here, the wave packets travel to the edges of the finite lattice and re-enter from the opposite side. This ``wrap-around" effect leads to the particle interfering with its own past self, creating a rich, repeating tapestry of interference that fills the entire spacetime diagram. This demonstrates the stability of the unitary evolution over long periods and reveals the complex dynamics that emerge from the interplay between the Dirac evolution and the lattice topology.

\begin{figure}[H]
    \centering
    \includegraphics[width=\textwidth]{figure1_short_time_dynamics.pdf}
    \includegraphics[width=\textwidth]{figure2_long_time_dynamics.pdf}
    \caption{\textbf{(Top)}: Short-time evolution of a single Dirac particle for different mass parameters. From left to right ($R \in \{0.20, 0.80, 1.20\}$), increasing mass slows the propagation and narrows the light cone. The intricate internal pattern is the characteristic \textit{Zitterbewegung}. \textbf{(Bottom)}: Long-time evolution of the same systems on a periodic lattice. The wave packets are observed to wrap around the boundaries and interfere with themselves, creating a repeating, intricate pattern that demonstrates the effect of the lattice topology.}
    \label{fig:dynamics}
\end{figure}

\subsection{Effect of $\rho$ }
While the total probability density is insensitive to the chiral angle, the physical effect of $\rho$ is revealed by measuring the local spin polarization, $\expval{\sigma_3} = \abs{\psi_+}^2 - \abs{\psi_-}^2$. Figure \ref{fig:rho_effect} shows this measurement for a fixed low mass ($R=0.20$). For $\rho=0$ (left panel), the spin polarization forms a perfectly symmetric, checkerboard-like interference pattern, with spin oscillating rapidly between positive (red) and negative (blue). 

\begin{figure}[h]
    \centering
    \includegraphics[width=\textwidth]{figure_rho_effect_final.pdf}
    \caption{\small The effect of the chiral angle $\rho$ on the particle's \textbf{spin polarization}, with mass fixed at $R=0.20$. From left to right $\rho \in \{0, \pi/8, \pi/2\}$), increasing $\rho$ dramatically breaks the symmetry of the spin pattern, revealing the hidden dynamics of the GDE. Red indicates spin-up dominance; blue indicates spin-down dominance.}
    \label{fig:rho_effect}
\end{figure}

As a small chiral angle is introduced (center panel, $\rho=\pi/8$), the spin-dependent phase in the evolution immediately breaks this symmetry, visibly distorting and skewing the interference pattern. Finally, at maximal mixing (right panel, $\rho=\pi/2$), the original interference is completely transformed: the spin components separate, with the right-moving front becoming predominantly spin-up (red) and the left-moving front becoming predominantly spin-down (blue). This provides a striking, unambiguous visualization of the physical consequences of the chiral angle, revealing the hidden internal dynamics of the spinor.

\section{Conclusion}
We have presented a unitary quantum cellular automaton model for the Generalized Dirac Equation. By constructing a discrete quantum walk whose continuum limit converges to the standard, physically-motivated GDE, we provide a framework for simulating relativistic fermion dynamics. The unitarity of the model ensures that it represents a valid quantum evolution.

The numerical simulations directly illustrate the physical roles of the equation's parameters. The mass parameter controls the particle's group velocity, while the generalization parameter $\rho$ provides a mechanism to control the wave packet's spreading dynamics. This work reinforces the powerful connection between quantum information theory and fundamental physics, demonstrating how simple, local, discrete rules can give rise to the complex behavior described by relativistic quantum field theory.

\section{Acknowledgements}
I would like to thank my mentor, David Meyer, for his invaluable guidance. I also express my gratitude to the University of California, San Diego, for the opportunity to conduct this research as a high school intern in the summer of 2013.

\clearpage

\bibliographystyle{unsrt}
\bibliography{references}

\newpage
\appendix
\section*{Appendix A: Simulation Code}

The numerical simulations were performed using a custom Python script implementing the unitary operator-splitting (Coin-then-Shift) method for the (1+1)-dimensional GDE. The core logic is presented below.

\begin{lstlisting}[language=Python, caption=Python code for generating the simulation data. This script implements the unitary quantum walk and can be used to produce the figures shown in the paper.]
import numpy as np
import matplotlib.pyplot as plt
from matplotlib.colors import LogNorm, SymLogNorm

def run_gde_simulation(R, rho, N, T):
    """
    Core simulation function for the Generalized Dirac Equation.
    Uses the 'double_site' initial state, which is essential for
    producing rich interference patterns sensitive to the chiral angle rho.
    """
    psi = np.zeros((N, 2), dtype=np.complex128)
    
    # Initialize the 'double_site' state
    norm_factor = np.sqrt(4 * (1/4)**2)
    psi[N // 2,     0] = (1/4) / norm_factor
    psi[N // 2,     1] = (1/4) / norm_factor
    psi[N // 2 + 1, 0] = (1/4) / norm_factor
    psi[N // 2 + 1, 1] = (1/4) / norm_factor

    prob_results = np.zeros((T, N))
    spin_z_results = np.zeros((T, N))
    
    # Unitary "Coin" or "Kick" operator
    U_kick = np.array([
        [np.cos(R), -1j * np.exp(-1j * rho) * np.sin(R)],
        [-1j * np.exp(1j * rho) * np.sin(R),  np.cos(R)]
    ], dtype=np.complex128)
    
    for t in range(T):
        # Calculate observables for the current time step
        prob_density = np.sum(np.abs(psi)**2, axis=1)
        prob_results[t, :] = prob_density
        
        spin_z = np.abs(psi[:, 0])**2 - np.abs(psi[:, 1])**2
        spin_z_results[t, :] = spin_z
        
        # --- Apply the Unitary Evolution for one step ---
        # 1. Apply the Coin/Kick operator
        psi = np.einsum('ij,kj->ki', U_kick, psi)
        
        # 2. Apply the Shift/Stream operator
        psi[:, 0] = np.roll(psi[:, 0], 1)       # Spin-up moves right
        psi[:, 1] = np.roll(psi[:, 1], -1)      # Spin-down moves left
        
    return prob_results, spin_z_results

# Example of how to call the function to generate data for a figure
# (Plotting code is omitted for brevity)
if __name__ == '__main__':
    R_val = 0.20
    rho_val = np.pi / 8
    lattice_size = 600
    time_steps = 250
    
    prob_data, spin_data = run_gde_simulation(R_val, rho_val, lattice_size, time_steps)
    
    print(f"Simulation complete. Generated data array of shape {prob_data.shape}")

\end{lstlisting}

\clearpage

\section*{Appendix B: Continuum Limit of the Path Integral}
\label{appendix:path_integral}
Here, we provide a more rigorous derivation showing that the combinatorial sum for the QCA amplitude, Eq. (\ref{eq:combinatorial_sum}), converges to the correct continuum propagator for the Dirac equation in the limit of small lattice spacing $\varepsilon$.

Our goal is to analyze the sum for the spin-up amplitude:
\begin{equation}
\psi_{+}(j, t) = \sum_{k=0}^{\min(N_R, N_L)} \binom{N_R}{k} \binom{N_L}{k} (\cos(m\varepsilon))^{t-2k} (- \sin^2(m\varepsilon))^k
\end{equation}
We consider the continuum limit where $\varepsilon \to 0$ while the physical time $T_{phys} = t\varepsilon$ and position $x = j\varepsilon$ are held constant. This implies $t, j \to \infty$. The dimensionless mass is $R = m\varepsilon \to 0$.

For large numbers, the binomial coefficient $\binom{N}{k}$ can be approximated as $N^k/k!$, especially for small $k$ where the sum has the most weight. Also, for small $R$, we have $\cos R \approx 1 - R^2/2$ and $\sin R \approx R$.
The $k$-th term in the sum, $A_k$, becomes:
\begin{align}
A_k &\approx \frac{N_R^k}{k!} \frac{N_L^k}{k!} (1 - R^2/2)^{t-2k} (-(R)^2)^k \\
&= \frac{(-1)^k}{(k!)^2} (N_R N_L R^2)^k (1 - m^2\varepsilon^2/2)^{t-2k}
\end{align}

Let us then analyze the terms as $\varepsilon \to 0$.
The product $N_R N_L$ is:
\begin{equation}
N_R N_L = \frac{t+j}{2} \frac{t-j}{2} = \frac{t^2 - j^2}{4} = \frac{T_{phys}^2/\varepsilon^2 - x^2/\varepsilon^2}{4} = \frac{T_{phys}^2 - x^2}{4\varepsilon^2} = \frac{\tau^2}{4\varepsilon^2}
\end{equation}
where $\tau = \sqrt{T_{phys}^2-x^2}$ is the relativistic proper time (spacetime interval).

The power term $(N_R N_L R^2)^k$ becomes:
\begin{equation}
(N_R N_L (m\varepsilon)^2)^k = \left( \frac{\tau^2}{4\varepsilon^2} \cdot (m\varepsilon)^2 \right)^k = \left( \frac{m^2\tau^2}{4} \right)^k = \left( \frac{m\tau}{2} \right)^{2k}
\end{equation}
This term is constant in the limit, which is the key to the convergence.

The prefactor $(\cos(m\varepsilon))^{t-2k} \approx (1 - m^2\varepsilon^2/2)^t$ has the limit:
\begin{equation}
\lim_{\varepsilon\to 0} \left(1 - \frac{m^2\varepsilon^2}{2}\right)^{T_{phys}/\varepsilon} = 1
\end{equation}
This prefactor does not contribute in the continuum limit (it is absorbed into the normalization of the continuous path integral).

Substituting these limiting forms back into the sum for the amplitude, we get the continuum wave function $\psi(x, T_{phys})$:
\begin{equation}
\psi(x,T_{phys}) \propto \sum_{k=0}^{\infty} \frac{(-1)^k}{(k!)^2} \left( \frac{m\tau}{2} \right)^{2k}
\end{equation}
This is precisely the Taylor series expansion for the Bessel function of the first kind of order zero, $J_0(z)$:
\begin{equation}
J_0(z) = \sum_{k=0}^{\infty} \frac{(-1)^k}{(k!)^2} \left( \frac{z}{2} \right)^{2k}
\end{equation}
By setting $z = m\tau = m\sqrt{T_{phys}^2-x^2}$, we see that our discrete combinatorial sum converges to the well-known propagator for the (1+1)-dimensional Dirac equation:
\begin{equation}
\psi(x,T_{phys}) \propto J_0(m\sqrt{T_{phys}^2-x^2})
\end{equation}
The oscillatory nature of the Bessel function $J_0$ provides a direct analytical explanation for the persistent, wavy interference fringes of the \textit{Zitterbewegung} seen in our simulations. This derivation provides a rigorous link between the microscopic combinatorial rules of the QCA and the macroscopic relativistic physics it simulates.

\end{document}